\documentclass[]{aa}
\usepackage[normalem]{ulem}
\usepackage{times}
\usepackage{mathptm}
\usepackage[dvips]{epsfig}
\usepackage[dvips]{rotating}
\usepackage{natbib}
\usepackage{times}
\bibpunct{(}{)}{;}{a}{}{,}

\newcommand{\cs}{CS~29497--004}
\newcommand{\tefft}{$T_{\mbox{\scriptsize eff}}$}
\newcommand{\teffm}{T_{\mbox{\scriptsize eff}}}
\newcommand{\rb}[1]{\raisebox{1.5ex}[-1.5ex]{#1}}

\begin{document}

\title{The Hamburg/ESO R-process Enhanced Star survey (HERES)\thanks{Based
in large part on observations collected at the European Southern Observatory, 
Paranal, Chile (proposal number 68.B-0320).}} 

\subtitle{I. Project description, and discovery of two stars with strong
 enhancements of neutron-capture elements}

\author{
  N. Christlieb\inst{1,3} \and 
  T. C. Beers\inst{2} \and 
  P. S. Barklem\inst{3} \and
  M. Bessell\inst{4} \and
  V. Hill\inst{5} \and
  J. Holmberg\inst{6,7} \and
  A.J. Korn\inst{3,8} \and
  B. Marsteller\inst{2} \and
  L. Mashonkina\inst{9,8} \and
  Y.-Z. Qian\inst{10} \and
  S. Rossi\inst{11} \and
  G.J. Wasserburg\inst{12} \and
  F.-J. Zickgraf\inst{1} \and
  K.-L. Kratz\inst{13,14} \and
  B. Nordstr{\"o}m\inst{5,15} \and
  B. Pfeiffer\inst{13,14} \and
  J. Rhee\inst{16,17}
  S.G. Ryan\inst{18}
}

\offprints{N. Christlieb,\\ \email{nchristlieb@hs.uni-hamburg.de}}

\institute{
     Hamburger Sternwarte, Universit\"at Hamburg, Gojenbergsweg 112,
     D-21029 Hamburg, Germany 
\and Department of Physics and Astronomy, Michigan State University, East
     Lansing, MI 48824, USA 
\and Department of Astronomy and Space Physics, Uppsala University, Box 515, 
     S-75120 Uppsala, Sweden 
\and Research School of Astronomy \& Astrophysics, Australian National
     University, Cotter Road, Weston, ACT 2611, Australia 
\and Observatoire de Paris, GEPI and URA 8111 du CNRS, 92195 Meudon Cedex,
     France 
\and Astronomical Observatory, NBIfAFG, Juliane Meries Vej 30, 2100
     Copenhagen, Denmark 
\and Nordic Optical Telescope Scientific Association, Apartado 474, ES-38\,700
     Santa Cruz de La Palma, Spain 
\and Universit\"ats-Sternwarte M\"unchen, Scheinerstrasse 1, D-81679 M\"unchen, 
     Germany 
\and Institute of Astronomy, Russian Academy of Sciences, Pyatnitskaya 48,
     109017 Moscow, Russia 
\and School of Physics and Astronomy, University of Minnesota, Minneapolis, 
     MN~55455, USA 
\and Departamento de Astronomia Instituto de Astronomia, Geof{\'\i}sica e 
     Ci{\^e}ncias Atmosf{\'e}ricas, Universidade de S{\~a}o Paulo, 05508-900 
     S{\~a}o Paulo SP, Brazil 
\and The Lunatic Asylum, Division of Geophysics and Planetary Sciences, 
     California Institute of Technology, Pasadena, CA91125, USA 
\and Institut f\"ur Kernchemie, Universit\"at Mainz, D-55128 Mainz, Germany 
\and VISTARS HGF -- Virtual Institute for Nuclear Structure and
     Nuclear Astrophysics, Mainz 
\and Lund University Box 43, S-22100 Lund, Sweden 
\and Center for Space Astrophysics, Yonsei University Seoul, 120-749, Korea 
\and Space Astrophysics Laboratory, California Institute of Technology, 
     MC 405-47, Pasadena, CA 91125, USA 
\and Department of Physics and Astronomy, Open University, Walton Hall, Milton
     Keynes, MK7~6AA, UK 
}

\date{Received  / Accepted }

\abstract{ 
We report on a dedicated effort to identify and study metal-poor stars
strongly enhanced in r-process elements ($\mbox{[r/Fe]}>1$\,dex; hereafter
r-II stars), the Hamburg/ESO R-process Enhanced Star survey
(HERES). Moderate-resolution ($\sim 2$\,{\AA}) follow-up spectroscopy has been
obtained for metal-poor giant candidates selected from the Hamburg/ESO
objective-prism survey (HES) as well as the HK survey to identify sharp-lined
stars with $\mbox{[Fe/H]}<-2.5$\,dex. For several hundred confirmed metal-poor
giants brighter than $B\sim 16.5$\,mag (most of them from the HES),
``snapshot'' spectra ($R\sim 20,000$; $S/N \sim 30$ per pixel) are being
obtained with VLT/UVES, with the main aim of finding the $2$--$3$\,\% r-II
stars expected to be among them. These are studied in detail by means of
higher resolution and higher $S/N$ spectra. In this paper we describe
a pilot study based on a set of 35 stars, including 23 from the HK survey, 8
from the HES, and 4 comparison stars. We discovered two new r-II stars, {\cs}
($\mbox{[Eu/Fe]}=1.64\pm 0.22$) and CS~29491-069 ($\mbox{[Eu/Fe]}=1.08\pm
0.23$). A first abundance analysis of {\cs} yields that its abundances of Ba
to Dy are on average enhanced by $1.5$\,dex with respect to iron and the Sun
and match a scaled solar r-process pattern well, while Th is underabundant
relative to that pattern by 0.3\,dex, which we attribute to radioactive
decay. That is, {\cs} seems not to belong to the class of r-process enhanced
stars displaying an ``actinide boost'', like CS~31082--001 (Hill et al. 2002),
or CS~30306--132 (Honda et al. 2004b).  The abundance pattern agrees well with
predictions of the phenomenological model of Qian \& Wasserburg.
\keywords{Surveys -- Stars:abundances -- Stars:population~II --
Galaxy:abundances -- Galaxy:halo} }

\titlerunning{HERES Project Description}
\authorrunning{Christlieb et al.}

\maketitle





\section{Introduction}

%
%
In the past few years a great deal of attention has become focused on the very
rare class of objects known as the ``r-process-enhanced metal-poor''
stars. These objects are enormously important as they allow to study, among
other things, the nature of the rapid neutron-capture process(es), and
possibly identify the site(s) for this nucleosynthesis process. Furthermore,
and perhaps even more importantly, individual age determinations are possible
for these stars using long-lived radioactive isotopes, such as
\element[][232]{Th} (half-life $14.05$\,Gyr) or \element[][238]{U}
($4.468$\,Gyr). By comparing the abundance ratio of these elements relative to
a stable r-process element of similar mass to the production ratio expected
from theoretical r-process yields, the time elapsed since the nucleosynthesis
event that produced these elements took place (e.g., in a type-II supernova)
can be derived. The time between this nucleosynthesis event and the birth of
stars that formed from gas clouds enriched by this material, including the
low-mass, r-process-enhanced stars that we observe today, is considered as
neglible compared the age of the star. Therefore, abundance ratios like Th/Eu,
U/Eu, or U/Th can be used as chronometers for age determination of
r-process-enhanced stars.

%
%

%
%
For ease of discussion, we divide the r-process enhancement phenomenon in
metal-poor stars into two categories:
\begin{description}
\item[\bf r-I:] Metal-poor stars with $+0.3 \le \mbox{[Eu/Fe]} \le +1.0$ and
   $\mbox{[Ba/Eu]} < 0$.
\item[\bf r-II:] Metal-poor stars with $\mbox{[Eu/Fe]} > +1.0$ and
   $\mbox{[Ba/Eu]} < 0$.
\end{description}
Note that the term 'metal-poor' is not necessarily refering to the overall
metal-content of the star, which might in fact not be significantly below the
solar value when the star under consideration also has strong overabundances
of C, N, and O. We use Eu as reference element for the neutron-capture
elements which were mainly produced by the r-process in the solar material,
because its abundance is most easily measurable. We include the condition
$\mbox{[Ba/Eu]} < 0$ into the above definitions, because Eu \emph{can} be
produced by the s-process as well. Therefore we need to distinguish between
``pure'' r-process-enhanced stars and stars that were enriched by material
produced in the r- \emph{and} s-process (like CS~22948$-$27 and CS~29497$-$34,
and HE~2148$-$1247; see \citealt{Hilletal:2000} and \citealt{Cohenetal:2003},
respectively), or only by the s-process.  Adopting the values of
\citet{Burrisetal:2000} for the solar Ba and Eu abundances and r- and
s-fractions of these elements, it follows that a star purely enriched by
neutron-capture elements produced in the s-process would have
$\mbox{[Ba/Eu]}_{\mbox{\small s}} = +1.5$, while $\mbox{[Ba/Eu]}_{\mbox{\small
r}} = -0.82$ for a pure r-process star.

It is not yet clear whether the division between r-I and r-II stars is
physically meaningful, in the sense that the r-I and r-II stars are associated
with different nucleosynthetic histories.  It is not even clear whether the
distribution of r-process-element enhancements, relative to iron, is best
described as bi-modal.  For now, this is only a division of convenience.

%
%
The first r-II star, CS~22892--052, was originally identified in the
compilation of metal-poor HK survey objects of Beers, Preston and Shectman
\citep{BPSII}. At that time it stood out because of its unusually strong CH
G-band for its low metallicity ($\mbox{[Fe/H]} = -3.1$). Soon afterwards, it
was observed at high resolution by \citet{McWilliametal:1995b}.  It was
immediately apparent that CS~22892--052 was clearly of interest for other
reasons as well -- the strength of absorption features associated with the
r-process, such as \ion{Eu}{ii}, were far greater than had been previously
observed in giants of such low metallicity.

\citet{Snedenetal:1996} showed that numerous other elements, including some
never detected before in metal-poor stars (such as Tb, Ho, Tm, Hf, and Os)
exhibited abundances, relative to iron, that were between 30 and 50 times
greater than observed in the Sun ($+1.2 \le \mbox{[r-process/Fe]} \le +1.6$),
making it a completely unique object at the time.  Of further interest, an
absorption line of Th was of measurable strength, which led to the use, by a
number of authors, of the Th/Eu chronometer to estimate the age of this star
\citep{Snedenetal:1996,Cowanetal:1997,Cowanetal:1999,Snedenetal:2003}.
This permitted, for the first time in an extremely low metallicity star, the
use of this technique to place a strong lower limit on the age of the Galaxy,
and hence of the Universe. \citet{Snedenetal:2003}, using new calculations for
the Th/Eu production ratio, determined an age of $12.8\pm 3$\,Gyr for
CS~22892--052. In the cosmological model fit to data recently obtained with
the Wilkinson Microwave Anisotropy Probe, and extensive redshift-survey data,
the age of the Universe is $13.7\pm 0.2$\,Gyr, which is consistent with the
lower limit following from the age of CS~22892--052.

%
%
Fortunately, CS~22892--052 is not unique.  During the course of a Large
Programme of Cayrel et al. (``First Stars'') carried out at the European
Southern Observatory (ESO), high-resolution observations of the very
metal-poor ($\mbox{[Fe/H]} = -2.9$) giant CS~31082--001 revealed it to be an
r-II star as well ($\mbox{[Eu/Fe]} = +1.6$). However, in contrast to
CS~22892--052, CS~31082--001 is not strongly carbon-enhanced, immediately
calling into question any causal connection between the enhancement of carbon
and the r-process enhancement phenomenon.

In CS~31082--001, it was possible for the first time in a metal-deficient star
to detect not only thorium, but also uranium, a potentially much more useful
chronometer than Th \citep{Cayreletal:2001}. The simultaneous detection of U
and Th allowed for the first use of the U/Th chronometer, which has less
reliance on the uncertain nuclear physics associated with the r-process than
does the Th/Eu chronometer \citep[e.g.,][]{Schatzetal:2002,Wanajoetal:2002}.
However, while [Eu/Fe] (and the ratios of other r-process elements relative to
iron) were enhanced by a similar factor as observed in CS~22892--052,
\citet{Hilletal:2002} noted that Th/Fe was almost a factor of 3 higher than
observed in CS~22892--052, i.e., $\log\left(\mbox{Th/Eu}\right) = -0.22$ as
compared to $\log\left(\mbox{Th/Eu}\right) = -0.62$ in CS~22892--052
\citep{Snedenetal:2003}. This fact makes questionable the use of chronometers
such as Th/Eu as well as other alternative chronometer pairs involving the
actinides Th or U in combination with lighter elements ($Z \le 70$). Recent
theoretical r-process calculations yield a production ratio of
$\log\left(\mbox{Th/Eu}\right)_0 = -0.35$ \citep{Snedenetal:2003}, and hence
it would follow that CS~31081--001 has a \emph{negative} age.

%
%
\citet{Hondaetal:2004a,Hondaetal:2004b} recently reported on Subaru/HDS
observations of 22 metal-poor stars with $\mbox{[Fe/H]} < -2.5$, including 8
stars from the HK survey which were not studied at high spectral resolution
before. They discovered one new r-I star (CS~30306--132;
$\mbox{[Eu/Fe]}=+0.85$), and one r-II star (CS~22183--031;
$\mbox{[Eu/Fe]}=+1.2$). CS~30306--132 is the second star known to be suffering
from an ``actinide boost'': its Th/Eu ratio of $\log\left(\mbox{Th/Eu}\right)
= -0.10$ is even higher than the ratio observed in
CS~31081--001. Unfortunately the Subaru/HDS spectrum of the r-II star
CS~22183--031 available to Honda et al. was not of sufficient quality to
detect Th.

%
%
Since the discovery of CS~22892--052, and the realization of the potential
power of the Th/Eu chronometer, a handful of r-I stars have been discussed in
the literature as well (e.g., HD~115444, \citealt{Westinetal:2000};
BD$+17^{\circ}\,3248$, \citealt{Cowanetal:2002}). These stars appear on the
whole to be more common than their extreme counterparts. They also provide
valuable information, not only through application of the Th/Eu chronometer to
obtain age limits, but on the remarkable consistency of the pattern of
r-process elements from star to star, which follows a scaled solar r-process
pattern within measurement uncertainties.

%
%

%
%
In conclusion, larger samples of r-process-enhanced metal-poor stars are
needed in order to make progress in our understanding of the phenomenon of
r-process enhancement, investigate how common the ``actinide boost''
phenomenon is and what its physical reasons might be, identification of
reliable chronometers for age determinations, and for constraining what the
possible site(s) of the r-process(es) are. However, there are several
fundamental problems which make the identification and study of these stars a
challenge. First of all, they are extremely rare. Based on high-resolution
spectroscopy that has been performed on the most metal-deficient giants, it
appears that r-II stars occur no more frequently than about 1 in 30 giants
with $\mbox{[Fe/H]} < -2.5$, i.e., roughly 3\,\%.  Due to the steep decrease
of the metallicity distribution of the galactic halo towards low metallicity
\citep{Ryan/Norris:1991,TimTSS}, giants in that [Fe/H] range are rare
themselves, although modern spectroscopic wide-angle surveys, like the HK and
Hamburg/ESO (HE) surveys, have succeeded in identifying such stars with
success rates as high as 10--20\,\% \citep{TimTSS,Christlieb:2003}. So, we are
faced with the daunting prospect of searching for a rare phenomenon amongst
rare objects. The r-I stars appear to be found with a frequency that is at
least a factor of two greater than this, and fortunately extend into the
higher metallicity stars (e.g., BD$+17^{\circ}\,3248$ with
$\mbox{[Fe/H]}=-2.1$), where a greater number of candidates exist.

Detection of uranium presents an even bigger challenge, due to the weakness of
the absorption lines involved, and blending with features of other species. It
was not possible to measure even the strongest uranium line in the optical,
\ion{U}{ii} 3859.57\,{\AA}, in the carbon-enhanced star CS~22892--052, because
of blending with a CN line. The U line is also close to a strong Fe line at
3859.9\,{\AA} \citep[see Fig. 10 of][]{Hilletal:2002}. The ideal star for
detecting uranium would therefore be a cool giant with low carbon abundance,
very low overall metallicity, but strong enhancement of the r-process
elements, and it should be a bright star, because high signal-to-noise ratios
($S/N$) as well as high spectral resolution ($R=\lambda/\Delta\lambda >
60,000$) are required to measure the \ion{U}{ii} 3859.57\,{\AA} line
accurately. Note that in CS~31082--001, this line has an equivalent width of
only $\sim 2$\,m{\AA}.

%
%
These reasons motivated us to start a systematic search for r-II stars. In
this paper we describe the approach we have chosen (Section
\ref{Sect:HERESapproach}). We present a sample of stars observed in a pilot
study (Section \ref{Sect:PilotStudySample}), and details of their observations
(Section \ref{Sect:Observations}). We report on the discovery of two new r-II
stars, CS~29497--004 ($\mbox{[Eu/Fe]}=1.64\pm 0.22$) and CS~29491--069
($\mbox{[Eu/Fe]}=1.08\pm 0.23$). A first abundance analysis of {\cs}, the more
strongly r-process enhanced star of the two, is described in Section
\ref{Sect:CS29497-004}, and in Section \ref{Sect:AbundancePattern} we discuss
its abundance pattern. We finish our paper with conclusions and remarks on
future prospects.

\section{The HERES approach}\label{Sect:HERESapproach}

\begin{figure}[htbp]
  \leavevmode
  \begin{center}
    \epsfig{file=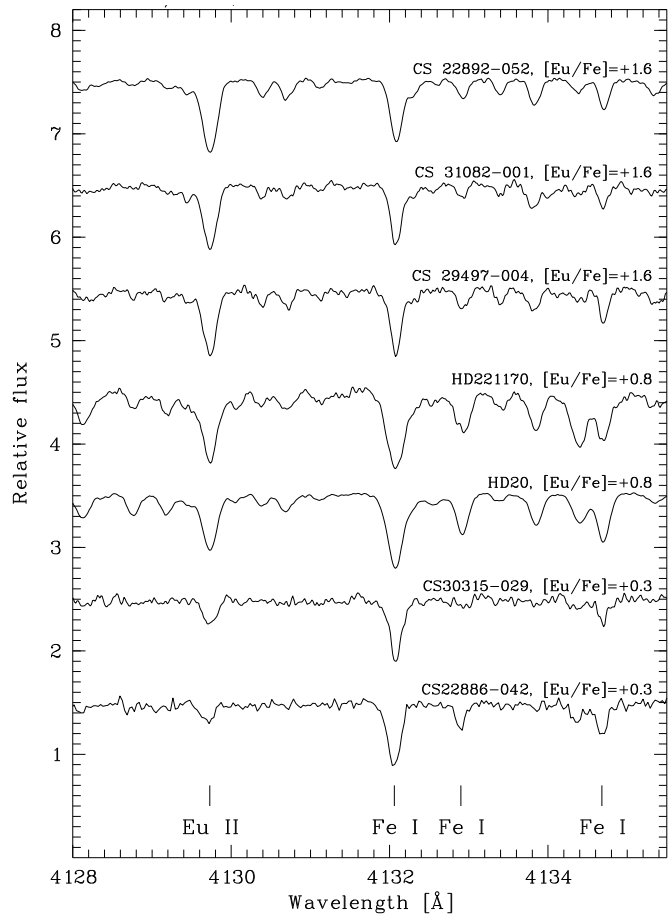, clip=, width=8.8cm, bbllx=41, bblly=130,
      bburx=483, bbury=740}
    \caption{\label{Eu4129demo_UVES} Spectral region around the \ion{Eu}{ii} 
       4129.73\,{\AA} line in several known r-I and r-II stars observed in
       our program for comparison reasons, and stars newly-identified in our 
       pilot study.}
  \end{center}
\end{figure}

For finding r-II stars, we adopt a two-step approach. The first step consists
of the identification of a large sample of metal-poor giants with
$\mbox{[Fe/H]}<-2.5$ in the Hamburg/ESO Survey (HES), by means of
moderate-resolution ($\sim 2$\,{\AA}) follow-up spectroscopy of several
thousand cool ($0.5 < B-V < 1.2$) metal-poor candidates selected in that
survey. In the second step, ``snapshot'' spectra ($S/N \sim 30$ per pixel at
$4100$\,{\AA}; $R\sim 20,000$) of confirmed metal-poor stars are
obtained. Such spectra can be secured for an $B=15.0$ star with a 8\,m-class
telescope in exposure times of only 900 seconds (see Table
\ref{Tab:Observations}), and under unfavourable observing conditions in terms
of seeing, fractional illumanation of and distance to the moon, cloud
coverage, and airmass. The weak constraints on the observing conditions makes
it feasible to observe large samples of stars.

As is demonstrated in Figure \ref{Eu4129demo_UVES}, snapshot spectra allow to
easily identify stars with enhancements of r-process elements, using the
\ion{Eu}{ii} 4129.73\,{\AA} line, since this line is very strong in these
stars. For example, in CS~22892--52, it has an equivalent width of
more than 100\,m{\AA}. 

We adopt the snapshot spectroscopy approach in a Large Programme (170.D-0010,
P.I. Christlieb) approved by ESO. A total of 373 stars (including 4 comparison
stars) are scheduled to be observed; most of them are from the HES.  These
observations are expected to yield $5$--$10$ new r-II stars, and at least twice
as many r-I stars.

As a by-product, our program will provide the opportunity to measure
abundances of $\alpha$-elements such as Mg, Ca, and Ti, iron-peak elements
such as Cr, Mn, Fe, Co, Ni, and Zn, as well as other elements, depending on
the $S/N$ of each spectrum, for the entire set of stars that we plan to
observe in snapshot mode. This will result in by far the largest sample of
very metal-poor stars with homogenously-measured abundances of a significant
number of individual elements.  Given the large number of spectra to be
processed, it is mandatory that we employ automated techniques for abundance
analysis. Such techniques are described in a companion paper
\citep[][hereafter Paper~II]{HERESpaperII}.

\section{The pilot study sample}\label{Sect:PilotStudySample}

\begin{table*}
  \centering
  \caption{\label{Tab:HERESsample} The pilot study sample. Coordinates of the
    HES stars have been derived from the Digitized Sky Survey~I and are
    accurate to $1''$; The coordinates of the HK survey stars are from
    identifications of the sources in the 2MASS All Sky Release. $V$
    magnitudes and $B-V$ colours have been measured with CCD photometry;
    [Fe/H] is derived from moderate-resolution follow-up spectroscopy using
    the methods of \citet{Beersetal:1999}. The photometry for HD~20 and
    HD~221170 has been retrieved from SIMBAD. Note that [Fe/H] as listed here
    was determined by using the most recent follow-up spectra as well as
    updates of the programs of \citet{Beersetal:1999}, therefore the
    metallicities have changed by a few tenths of a dex in some cases.  The
    best available [Fe/H] for the stars with $\mbox{[Fe/H]}>-2.5$ in the table
    was below $-2.5$\,dex when the target selection was done.  }
  \begin{tabular}{lrrrrrl}\hline
    Star & \multicolumn{1}{l}{$\alpha$ (2000.0)} & \multicolumn{1}{l}{$\delta$ (2000.0)} & 
    \multicolumn{1}{l}{$V$} & \multicolumn{1}{l}{$B-V$} & \multicolumn{1}{l}{[Fe/H]} & 
    Note \rule{0.0ex}{2.3ex}\\\hline
    CS~22175--007  & 02 17 26.6 & $-$09 00 45 & 13.461 & 0.676 & $-2.63$ & \rule{0.0ex}{2.3ex}\\
    CS~22186--023  & 04 19 45.5 & $-$36 51 35 & 12.841 & 0.682 & $-2.72$ & \\
    CS~22186--025  & 04 24 32.8 & $-$37 09 02 & 14.228 & 0.737 & $-2.77$ & Cayrel et al. LP\\
    CS~22886--042  & 22 20 25.8 & $-$10 23 20 & 13.263 & 0.831 & $-2.53$ & \\
    CS~22892--052  & 22 17 01.6 & $-$16 39 27 & 13.213 & 0.800 & $-2.97$ & r-II (comparison star)\\
    CS~22945--028  & 23 31 13.5 & $-$66 29 57 & 14.624 & 0.653 & $-2.38$ & \\
    CS~22957--013  & 23 55 49.0 & $-$05 22 52 & 14.089 & 0.775 & $-2.86$ & \\
    CS~22958--083  & 02 15 42.7 & $-$53 59 56 & 14.423 & 0.664 & $-2.68$ & \\
    CS~22960--010  & 22 08 25.3 & $-$44 53 56 & 13.944 & 0.497 & $-2.15$ & \\
    CS~29491--069  & 22 31 02.1 & $-$32 38 36 & 13.075 & 0.600 & $-2.81$ & r-II\\
    CS~29491--109  & 22 25 01.1 & $-$32 14 41 & 13.181 & 0.802 & $-2.37$ & \\
    CS~29497--004  & 00 28 06.9 & $-$26 03 04 & 14.034 & 0.705 & $-2.66$ & r-II\\
    CS~29510--058  & 02 21 46.5 & $-$24 01 58 & 13.510 & 0.669 & $-2.62$ & \\
    CS~30308--035  & 20 45 54.1 & $-$44 50 29 & 13.850 & 0.788 & $-3.23$ & \\
    CS~30315--001  & 23 37 38.8 & $-$26 21 53 & 13.763 & 0.902 & $-2.73$ & \\
    CS~30315--029  & 23 34 26.6 & $-$26 42 14 & 13.661 & 0.915 & $-3.29$ & r-I; Cayrel et al. LP\\
    CS~30337--097  & 22 01 21.5 & $-$30 57 57 & 13.202 & 0.722 & $-2.65$ & Cayrel et al. LP\\
    CS~30339--041  & 00 23 12.9 & $-$37 01 26 & 13.915 & 0.555 & $-2.42$ & \\
    CS~30343--063  & 21 45 17.5 & $-$37 22 18 & 13.008 & 1.045 & $-2.58$ & \\
    CS~31060--047  & 00 08 07.8 & $-$15 54 03 & 13.823 & 0.854 & $-2.87$ & \\
    CS~31062--041  & 00 35 03.0 & $-$15 54 29 & 13.934 & 0.820 & $-3.02$ & \\
    CS~31072--118  & 05 08 53.5 & $-$59 18 21 & 12.695 & 0.920 & $-2.94$ & \\
    CS~31082--001  & 01 29 31.1 & $-$16 00 45 & 11.666 & 0.766 & $-2.80$ & r-II (comparison star)\\
    HD~20          & 00 05 15.3 & $-$27 16 18 &   9.07 & 0.54  & $-1.25$ & r-I (comparison star)\\
    HD~221170      & 23 29 28.8 & $+$30 25 57 &   7.71 & 1.02  & $-1.61$ & r-I (comparison star)\\
    HE~0109$-$3711 & 01 11 38.4 & $-$36 55 17 & 16.290 & 0.378 & $-3.00$ & \\ 
    HE~0131$-$2740 & 01 33 25.8 & $-$27 25 28 & 14.632 & 0.590 & $-2.70$ & \\ 
    HE~0249$-$0126 & 02 51 39.7 & $-$01 14 33 & 15.685 & 0.571 & $-2.52$ & \\
    HE~0256$-$1109 & 02 59 10.1 & $-$10 58 01 & 15.608 & 0.549 & $-2.75$ & \\ 
    HE~0447$-$4858 & 04 49 01.0 & $-$48 53 35 & 16.254 & 0.433 & $-2.03$ & \\ 
    HE~0501$-$5139 & 05 02 48.1 & $-$51 35 36 & 16.094 & 0.479 & $-2.86$ & \\  
    HE~0513$-$4557 & 05 15 12.1 & $-$45 54 10 & 15.743 & 0.536 & $-3.00$ & \\ 
    HE~0519$-$5525 & 05 19 59.1 & $-$55 22 41 & 15.034 & 0.536 & $-2.42$ & \\ 
    HE~2145$-$3025 & 21 48 43.3 & $-$30 11 08 & 14.879 & 0.533 & $-2.40$ & \\ 
    HE~2338$-$1311 & 23 41 08.3 & $-$12 55 10 & 15.622 & 0.597 & $-2.85$ & \\\hline
  \end{tabular}
\end{table*}    

At the end of 2001, i.e., before the ESO Large Programme was approved, a
snapshot spectroscopy pilot study was conducted with the Ultraviolet-Visual
{\'E}chelle Spectrograph \citep[UVES;][]{Dekkeretal:2000} mounted on the 8\,m
Unit Telescope 2 (Kueyen) of the Very Large Telescope (VLT), in order to
verify the feasibility of our approach to find r-II stars. The targets were
drawn from lists of confirmed metal-poor stars with $\mbox{[Fe/H]}<-2.5$ from
the HK and Hamburg/ESO surveys, where [Fe/H] is determined from
moderate-resolution ($\sim 2$\,{\AA}) follow-up spectroscopy using the Ca~II~K
technique of \cite{Beersetal:1999}. We restricted the sample to stars with
$B-V > 0.5$, because we are primarily interested in cool, sharp-lined
giants. The $B-V$ colours are based on CCD photometry in case of the HK-survey
stars (recognisable by designations beginning with 'CS' for Curtis
Schmidt-telescope), or were derived directly from the HES objective-prism
spectra, in case of the HES stars (designations beginning with
'HE').\footnote{The nomenclature adopted in this and the original works
on CS stars has been modified when the stars were subsequently incorporated
into the SIMBAD database. For reasons associated with the uniqueness of SIMBAD
entries, stars which in the literature have designations CS~XXXXX-XXX are
listed in SIMBAD as BPS~CS~XXXXX-XXXX.} Despite of this colour selection, 2
stars later turned out to be dwarfs and a few to be subgiants hotter than
desired (for details see Table 4 of Paper~II). The main reason for this is
that the HES $B-V$ colours are of limited accuracy, i.e., $\sigma = 0.1$\,mag
\citep{HESStarsI}.

35 stars successfully observed with VLT/UVES for which CCD $BVRI$ photometry
is available to us form the sample of our pilot study.\footnote{For
completeness we mention that one additional star has been removed from the
sample because we found it to have a surface gravity of $\log g = 4.9$, which is
not covered by our grid of model atmospheres described in Paper~II, and it
appeared to be a fast rotator, making an automated abundance analysis
difficult.}  The sample is presented in Table \ref{Tab:HERESsample}. It
includes 4 comparison stars which are known to be r-I or r-II stars: HD~20
and HD~221170 \citep{Burrisetal:2000}, CS~22892--052
\citep{Snedenetal:1996,Snedenetal:2000,Snedenetal:2003}, and CS~31082--001
\citep{Cayreletal:2001,Hilletal:2002}. Three stars, CS~22186--025,
CS~30315--029, and CS~30337--097, are included in the ESO Large Programme
``The First Stars'' of Cayrel et al. Abundances from C to Zn for
CS~22186--025 have been published in \citet[][see that paper for further
references for the ``First Stars'' project]{Cayreletal:2004}.

Based on the results presented in Paper~II as well as Section
\ref{Sect:CS29497-004} below, we classify one of the 31 previously unobserved
stars in our sample as an r-I star (CS~30315--029), and two as r-II stars
(CS~29497--004\footnote{CCD photometry for this star has been previously
reported by \citet{Norrisetal:1999}.} and CS~29491-069).

\section{Observations}\label{Sect:Observations}

\begin{table}
  \centering
  \caption{\label{Tab:Observations} Observations of the pilot study
  sample. UT is the Universal Time at the start of the observations; 
  $S/N$ is the average signal-to-noise ratio per rebinned pixel in
  order \#113 ($\lambda = 4112.3\mbox{--}4148.9$\,{\AA}).
    }
  \begin{tabular}{lllrr}\hline
    Star & Date & UT & \multicolumn{1}{l}{$t_{\mbox{\tiny exp}}$} & 
    \multicolumn{1}{l}{$S/N$}\rule{0.0ex}{2.3ex}\\
        & & & \multicolumn{1}{l}{[sec]} & \\\hline
    CS~22175--007  & 2001-11-23 & 00:50:07 &  600 &  28\rule{0.0ex}{2.3ex}\\
    CS~22186--023  & 2001-10-10 & 08:50:08 &  600 &  51\\
    CS~22186--025  & 2001-10-08 & 08:52:52 &  900 &  29\\
    CS~22886--042  & 2001-11-29 & 00:50:15 &  600 &  36\\
    CS~22892--052  & 2001-11-01 & 00:33:42 &  600 &  41\\
    CS~22945--028  & 2001-12-02 & 02:04:55 &  900 &  27\\
    CS~22957--013  & 2001-11-22 & 02:03:40 &  900 &  31\\
    CS~22958--083  & 2001-11-01 & 03:44:10 &  900 &  29\\
    CS~22960--010  & 2001-10-07 & 03:08:57 &  900 &  32\\
    CS~29491--069  & 2001-12-02 & 01:49:30 &  600 &  51\\
    CS~29491--109  & 2001-12-02 & 01:32:46 &  600 &  44\\
    CS~29497--004  & 2001-11-01 & 03:04:07 &  900 &  35\\
    CS~29510--058  & 2001-11-23 & 05:43:00 &  600 &  34\\
    CS~30308--035  & 2001-11-01 & 00:02:09 &  600 &  29\\
    CS~30315--001  & 2001-12-02 & 02:25:14 &  600 &  32\\
    CS~30315--029  & 2001-12-03 & 00:45:57 &  600 &  33\\
    CS~30337--097  & 2001-10-07 & 02:53:24 &  600 &  40\\
    CS~30339--041  & 2001-11-01 & 02:45:56 &  600 &  34\\
    CS~30343--063  & 2001-10-07 & 02:37:17 &  600 &  34\\
    CS~31060--047  & 2001-11-22 & 00:42:29 &  600 &  26\\
    CS~31062--041  & 2001-11-01 & 03:24:03 &  900 &  33\\
    CS~31072--118  & 2001-10-09 & 08:47:15 &  600 &  47\\
    CS~31082--001  & 2001-11-22 & 01:01:11 &  600 &  81\\
    HD~20          & 2001-12-04 & 00:32:44 &   60 & 101\\
    HD~221170      & 2001-11-22 & 00:33:55 &   10 &  40\\
    HE~0109$-$3711 & 2001-11-23 & 03:05:57 & 1800 &  16\\
    HE~0131$-$2740 & 2001-11-23 & 02:37:45 & 1200 &  26\\
    HE~0249$-$0126 & 2001-11-23 & 01:34:02 & 1200 &  12\\
    HE~0256$-$1109 & 2001-11-23 & 02:00:28 & 1200 &  15\\
    HE~0447$-$4858 & 2001-11-23 & 03:42:26 & 1800 &  16\\
    HE~0501$-$5139 & 2001-11-13 & 07:53:40 & 1800 &  20\\
    HE~0513$-$4557 & 2001-11-12 & 05:18:27 & 1200 &  18\\
    HE~0519$-$5525 & 2001-11-17 & 07:53:23 & 1200 &  24\\
    HE~2145$-$3025 & 2001-11-01 & 00:48:50 &  900 &  24\\
    HE~2338$-$1311 & 2001-11-01 & 01:08:34 & 1200 &  19\\\hline
  \end{tabular}
\end{table}

Moderate-resolution spectroscopic follow-up observations were carried out in
numerous campaigns with a variety of telescope/instrument combinations,
including the SSO~2.3\,m/DBS, ESO~3.6\,m/EFOSC2, ESO~NTT/EMMI, ESO~1.5\,m/B\&C
Spectrograph, McDonald 2.7\,m/LCS, the R-C spectrographs mounted at the
KPNO~4\,m, KPNO~2.1\,m, CTIO 4\,m and CTIO~1.5\,m telescopes.

The VLT/UVES observations of the pilot study sample were carried out between 7
October and 4 December 2001. We used the BLUE~437 setup, yielding a wavelength
coverage of $\lambda=3760$--$4980$\,{\AA}. A $2''$ slit was chosen, typically
yielding (seeing-limited) resolving powers of $R=20,000$--$25,000$.  While
acquiring the data, the CCD binning was set to $1\times 1$ pixels, which
results in oversampling in combination with the wide slit. Therefore, we
rebinned the pipeline-reduced spectra by a factor of 2. The exposure times
were set such that a minimum nominal $S/N$ per rebinned pixel of $20$ was
reached, assuming a seeing of $2''$, full moon, thin cirrus, and $\sec z =
2$. However, given that the overhead for telescope pointing, image analysis,
target acquisition, CCD readout, etc., is $\sim 10$\,min, the minimum exposure
time was set to 10\,min (except for the very bright stars HD~20 and HD~221170,
which would saturate the detector in this exposure time) to avoid ineffecient
use of the telescope time. This resulted in $S/N>50$ for some of the brighter
stars. Furthermore, if the observing conditions were better than assumed in
our exposure time calculations, higher $S/N$ spectra were acquired. The
observations are summarized in Table \ref{Tab:Observations}.

CCD $BVRI$ photometry was obtained at the ESO-Danish 1.54\,m-telescope with
DFOSC. The observers were Beers \& Rossi (runs in December 1998 and 2000),
Holmberg (October 2002 and April 2003 runs) and Zickgraf (October 2003 run).
The data was reduced using the usual IRAF photometry reduction packages. The
stars were calibrated to the $BVR_CI_C$ system, (where the subscript 'C'
indicates Kron-Cousins) using a selection of stars from nightly observations
of Landolt standard fields. Typical errors in the resulting magnitudes are of
the order of $0.01$--$0.02$\,mag, i.e., better than the uncertainties expected
to arise from reddening corrections.  For further details, see Beers et
al. (2004, in preparation).  $V$ magnitudes and $B-V$ colours are listed in
Table \ref{Tab:HERESsample}; additional colours can be found in Table 3 of
Paper~II.

\section{Abundance analysis of CS~29497--004, a new r-II star}\label{Sect:CS29497-004}

A quick-look analysis of the snapshot spectrum of CS~29497--004, based on the
equivalent ratio of the \ion{Eu}{ii} 4129 and \ion{Fe}{i} 4132 lines
immediately revealed that this star is strongly enhanced in neutron-capture
elements. We therefore carried out a first abundance analysis. A more
comprehensive analysis, based on higher resolution and $S/N$ data already
obtained with VLT/UVES, will be presented in a forthcoming paper (Hill et
al. 2004, in preparation).

For the abundance analysis we used spherically symmetric model atmospheres
from a new grid of MARCS models \citep{Gustafssonetal:2004}. Enhancement of
$\alpha$-elements by $+0.4$\,dex has been taken into account.

\subsection{Stellar parameters}

\begin{table}
  \centering
  \caption{\label{Tab:Teff} Stellar parameters adopted
    for CS~29497-004.}
  \begin{tabular}{lll}\hline
    Colour & Value & {\tefft}\rule{0.0ex}{2.3ex}\\\hline
    $(V-J)_0$ & $1.539\pm 0.029$ & $5035 \pm 75$\,K \rule{0.0ex}{2.3ex}\\
    $(V-K)_0$ & $2.066\pm 0.032$ & $5110 \pm 50$\,K\\\hline
  \end{tabular}
\end{table}

\begin{table}
  \centering
  \caption{\label{Tab:StellarParameters} Stellar parameters adopted
    for CS~29497-004.}
  \begin{tabular}{lrr}\hline
    Parameter & \multicolumn{1}{l}{Value} & 
    \multicolumn{1}{l}{$\sigma$}\rule{0.0ex}{2.3ex}\\\hline
    $T_{\mbox{\scriptsize eff}}$ & \multicolumn{1}{l}{$5090$\,K}  & 
                                   \multicolumn{1}{l}{$100$\,K}\rule{0.0ex}{2.3ex}\\
    $\log g$ (cgs)                             & $2.4$\,dex          & $0.4$\,dex\\
    $\mbox{[Fe/H]}_{\mbox{\scriptsize LTE}}$   & $-2.64$\,dex        & $0.12$\,dex\\
    $v_{\mbox{\scriptsize micr}}$              & $1.6$\,km\,s$^{-1}$ & $0.2$\,km\,s$^{-1}$\\\hline
  \end{tabular}
\end{table}

The effective temperature {\tefft} was derived from broad-band visual ($V$)
and infrared ($JK$) photometry with the methods described in
\citet{KeckpaperI}. The $JK$ measurements are from the Two Micron All Sky
Survey\footnote{2MASS is a joint project of the University of Massachusetts
and the Infrared Processing and Analysis Center/California Institute of
Technology, funded by the National Aeronautics and Space Administration and
the U.S. National Science Foundation.} \citep[2MASS;][]{Skrutskieetal:1997}.
From the maps of \citet{Schlegeletal:1998} we computed $E(B-V)=0.016$ for the
interstellar reddening along the line of sight to CS~29497--004, and we
de-reddened the $V-J$ and $V-K$ colours using this value. The results are
listed in Table \ref{Tab:Teff}.

We adopt the weighted average of the effective temperatures deduced from these
colours, $\teffm = 5090$\,K. The formal error for {\tefft} is small, i.e., of
the order of $50$\,K, but a comparison with values deduced from Balmer line
profile analysis as well as the photometrically determined value of 5013\,K
reported in Paper~II reveals that systematic errors might be larger, of the
order of $\sim 100$\,K.

Surface gravities $\log g$ can be determined from the \ion{Fe}{i}/\ion{Fe}{ii}
ionisation equilibrium. \ion{Fe}{i} is, however, a minority species in cool
giants and therefore potentially subject to formation under non-LTE conditions
which could affect gravity estimates in a systematic way. \ion{Fe}{i} non-LTE
calculations for metal-poor stars were recently carried out by
\citet{Kornetal:2003}. Applying these calculations (updated with respect to
the treatment of Rayleigh scattering) to {\cs}, we find non-LTE corrections
for the \ion{Fe}{i} abundance of $+0.10$\,dex, corresponding to a correction
in $\log g$ of $+0.25$\,dex. An analysis differentially to the Sun performed
by A. Korn with a plane-parallel MAFAGS model
\citep{Gehren:1975a,Gehren:1975b} led to a non-LTE gravity of $\log g = 2.25$,
where both \ion{Fe}{i} and \ion{Fe}{ii} give [Fe/H] = $-2.63\pm
0.09$. Inspection of a 12\,Gyr isochrone for $Z = 0.00001\,Z_{\odot}$
\citep{Yietal:2001} reveals that a star of $\teffm = 5090$\,K on the giant
branch has $\log g=2.28$, in very good agreement with the non-LTE
analysis. The non-differential analysis resulted in $\log g=2.35$. The
determination by A. Korn was limited to 14 \ion{Fe}{i} and 7 \ion{Fe}{ii}
lines in the yellow spectral region which were chosen based on their
relatively undisturbed line profiles in the solar spectrum. Many more
\ion{Fe}{i} lines are available in the blue part of the spectrum of {\cs}
(heavily blended at solar metallicity). Using this extended line list with
$\log gf$ values from the sources listed in Tables 7--9 of Paper~II and
spherically-symmetric MARCS models, the \ion{Fe}{i}/\ion{Fe}{ii} ionisation
equilibrium is reached at $\log g = 2.4$. Applying the non-LTE correction of
$+0.25$\,dex to this value would result in an unrealistically high gravity of
$\log g = 2.65$, as judged by the comparison with isochrones. Therefore, and
for self-consistency reasons, we adopt the non-differential LTE gravity of
$\log g = 2.4$ derived with the MARCS models we also use in our abundance
analysis. The formal uncertainty, derived from the uncertainties of the
abundances of \ion{Fe}{i} and \ion{Fe}{ii} (see Table \ref{Tab:Abundances}),
is $\Delta\log g = 0.4$\,dex.

The LTE iron abundance we derive from the UVES snapshot spectra is
$\mbox{[Fe/H]} = -2.64\pm 0.12$, which agrees very well with the value of
$\mbox{[Fe/H]} = -2.66\pm 0.3$ we found from moderate-resolution follow-up
spectroscopy (see Table \ref{Tab:HERESsample}).

The stellar parameters adopted for our first abundance analysis of {\cs} are
summarized in Table \ref{Tab:StellarParameters}.

\subsection{Abundances of carbon to dysprosium}

\begin{figure}[htbp]
  \leavevmode
  \begin{center}
    \epsfig{file=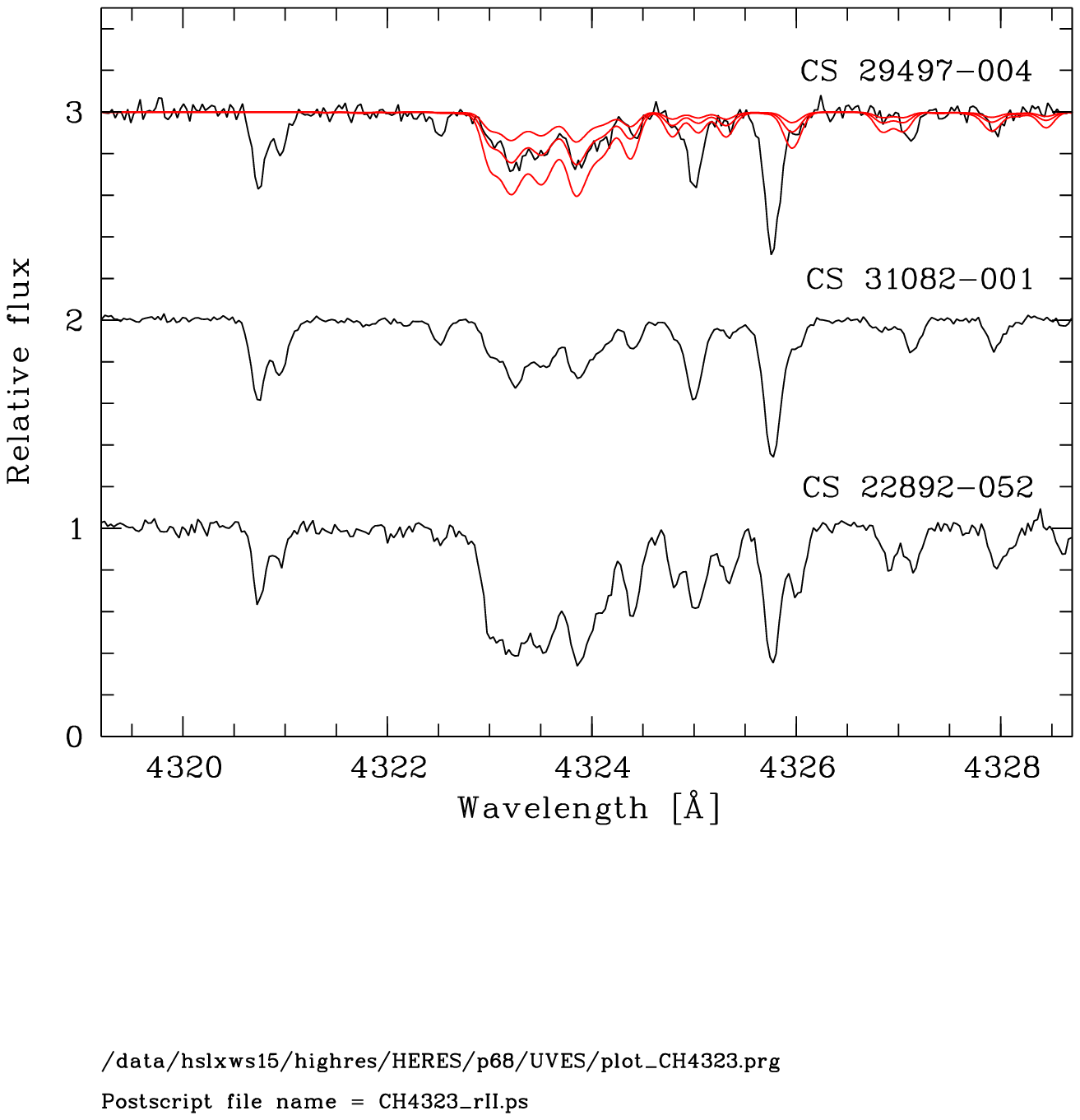, clip=, width=8.8cm, 
            bbllx=76, bblly=278, bburx=455, bbury=571}
    \caption{\label{Fig:CHfit} The CH feature at $\sim 4323$\,{\AA} in {\cs},
             CS~22892--052 and CS~31082--001. For {\cs}, we show a comparison
             with synthetic spectra computed with $\log \epsilon (\mbox{C}) = 
             5.56$, $5.86$, and $6.16$. }
  \end{center}
\end{figure}

From the UVES snapshot spectra of {\cs} we derive abundances for 18 elements,
including 11 neutron-capture elements. For most of them, we measured
equivalent widths of clean lines by fitting simultaneously a Gaussian profile
and a local straight-line continuum. We used the line list presented in
Paper~II, with the exception of the lines employed in spectrum synthesis
calculations for the wavelength region around the \ion{Th}{ii} 4019\,{\AA}
line (see Section \ref{Sect:Th4019_synthesis} below).

Hyperfine-structure (HFS) splitting has been taken into account for Sr, Ba and
Eu (for sources of the HFS calculations and $\log gf$ values of the lines we
used see Paper~II). We assume a pure r-process isotopic mixture for Ba and Eu,
and solar isotopic ratios for Sr. The abundances of these elements have been
determined by spectrum synthesis of the following lines: \ion{Ba}{ii}
4554.00\,{\AA}, \ion{Ba}{ii} 4934.10\,{\AA}, \ion{Eu}{ii} 3819.67\,{\AA},
\ion{Eu}{ii} 4129.72\,{\AA}, and \ion{Sr}{ii} 4215.54\,{\AA}.  We note that
the Sr and Ba abundances have a considerable sensitivity against changes in
the microturbulence. Varying $v_{\mbox{\tiny micr}}$ by $0.2$\,km/s results in
changes of the Ba and Sr abundance (as determined from the lines at
4215\,{\AA} and 4554\,{\AA}, respectively) of $0.1$\,dex.  In contrast, the
effect on the \ion{Eu}{ii} 4129.72\,{\AA} is very small, i.e.,
$\Delta\log\epsilon/\Delta v_{\mbox{\tiny micr}} =
0.01\,\mbox{dex}/0.2$\,km/s.

The carbon abundance of {\cs} was measured from CH features around
$4323$\,{\AA}. In Figure \ref{Fig:CHfit} we show a comparison of that spectral
region in {\cs} with that of CS~22892--052 and CS~31082--001. The CH features
in {\cs} are much weaker than in CS~22892--052 and about as weak as in
CS~31082--001, making {\cs} a good candidate for an attempt to detect
uranium. We derived $\log \epsilon (\mbox{C}) = 5.86$ for {\cs}. For
comparison, the C abundances of CS~22892--052 and CS~31082--001 are $\log
\epsilon (\mbox{C}) = 6.30$ \citep{Snedenetal:2003} and $\log \epsilon
(\mbox{C}) = 5.82$ \citep{Hilletal:2002}, respectively.
 
From the absence of \element[][13]{CH} features in the spectral region around
4224\,{\AA} as well as the best fit we could achieve in the wavelength region
around the \ion{Th}{ii} 4019\,{\AA} line (see below), we derive a lower limit
for the \element[][12]{C}/\element[][13]{C} ratio of $10$.

\begin{table}
  \centering
  \caption{\label{Tab:Abundances} Abundances of {\cs}. $\sigma$ refers to 
     the 1-$\sigma$ line-to-line scatter in case of equivalent-width
     based analysis. In case of the spectrum synthesis results, errors
     were roughly estimated from the sensitivity of the line strengths
     to changes of the abundances. In the computation of [X/Fe], we 
     compared the abundances of neutral species with
     $\log\epsilon\left(\mbox{Fe~{\sc i}}\right)$, and the abundances of
     singly ionized species with $\log\epsilon\left(\mbox{Fe~{\sc
     ii}}\right)$. 
    }
  \begin{tabular}{lrrrrrr}\hline
    Species & \multicolumn{1}{l}{$N_{\mbox{\tiny lines}}$} &
    \multicolumn{1}{l}{$\log \epsilon$} &
    \multicolumn{1}{l}{$\sigma_{\log \epsilon}$} &
    \multicolumn{1}{l}{$(\log \epsilon)_{\odot}$} &
    \multicolumn{1}{l}{[X/H]} &
    \multicolumn{1}{l}{[X/Fe]} \rule{0.0ex}{2.3ex} \\\hline
    CH           & syn     & $ 5.86$ & $0.10$ & 8.39 & $-2.53$ & $+0.10$\rule{0.0ex}{2.3ex}\\
    \ion{Mg}{i}  &      4  & $ 5.21$ & $0.14$ & 7.58 & $-2.37$ & $+0.26$\\
    \ion{Ca}{i}  &      5  & $ 4.04$ & $0.17$ & 6.36 & $-2.32$ & $+0.31$\\
    \ion{Ti}{i}  &      3  & $ 2.46$ & $0.07$ & 4.99 & $-2.53$ & $+0.10$\\
    \ion{Ti}{ii} &      8  & $ 2.67$ & $0.15$ & 4.99 & $-2.32$ & $+0.32$\\
    \ion{V}{i}   &      1  & $ 1.37$ & $    $ & 4.00 & $-2.63$ & $ 0.00$\\
    \ion{Fe}{i}  &     28  & $ 4.82$ & $0.16$ & 7.45 & $-2.63$ & $$\\
    \ion{Fe}{ii} &      7  & $ 4.81$ & $0.12$ & 7.45 & $-2.64$ & $$\\
    \ion{Zn}{i}  &      1  & $ 2.20$ & $    $ & 4.60 & $-2.40$ & $+0.23$\\
    \ion{Sr}{ii} & syn (1) & $ 0.88$ & $0.20$ & 2.97 & $-2.09$ & $+0.55$\\
    \ion{Y}{ii}  &      4  & $ 0.30$ & $0.06$ & 2.24 & $-1.94$ & $+0.70$\\
    \ion{Zr}{ii} &      2  & $ 0.98$ & $    $ & 2.60 & $-1.62$ & $+1.02$\\
    \ion{Ba}{ii} & syn (2) & $ 0.50$ & $0.20$ & 2.13 & $-1.63$ & $+1.01$\\
    \ion{Ce}{ii} &      4  & $ 0.07$ & $0.05$ & 1.58 & $-1.51$ & $+1.13$\\
    \ion{Nd}{ii} &      9  & $ 0.23$ & $0.12$ & 1.50 & $-1.27$ & $+1.37$\\
    \ion{Sm}{ii} &      2  & $ 0.05$ & $    $ & 1.01 & $-0.96$ & $+1.68$\\
    \ion{Eu}{ii} & syn (2) & $-0.45$ & $0.20$ & 0.51 & $-0.96$ & $+1.68$\\
    \ion{Gd}{ii} &      1  & $ 0.27$ & $    $ & 1.12 & $-0.85$ & $+1.79$\\
    \ion{Dy}{ii} &      1  & $ 0.38$ & $    $ & 1.14 & $-0.76$ & $+1.88$\\
    \ion{Th}{ii} & syn (1) & $-0.96$ & $0.15$ & 0.12 & $-1.08$ & $+1.56$\\\hline
   \end{tabular}
\end{table}

\subsection{Spectrum synthesis of the \ion{Th}{ii} 4019\,{\AA}
  region}\label{Sect:Th4019_synthesis} 

\begin{figure*}[htbp]
  \leavevmode
  \begin{center}
    \epsfig{file=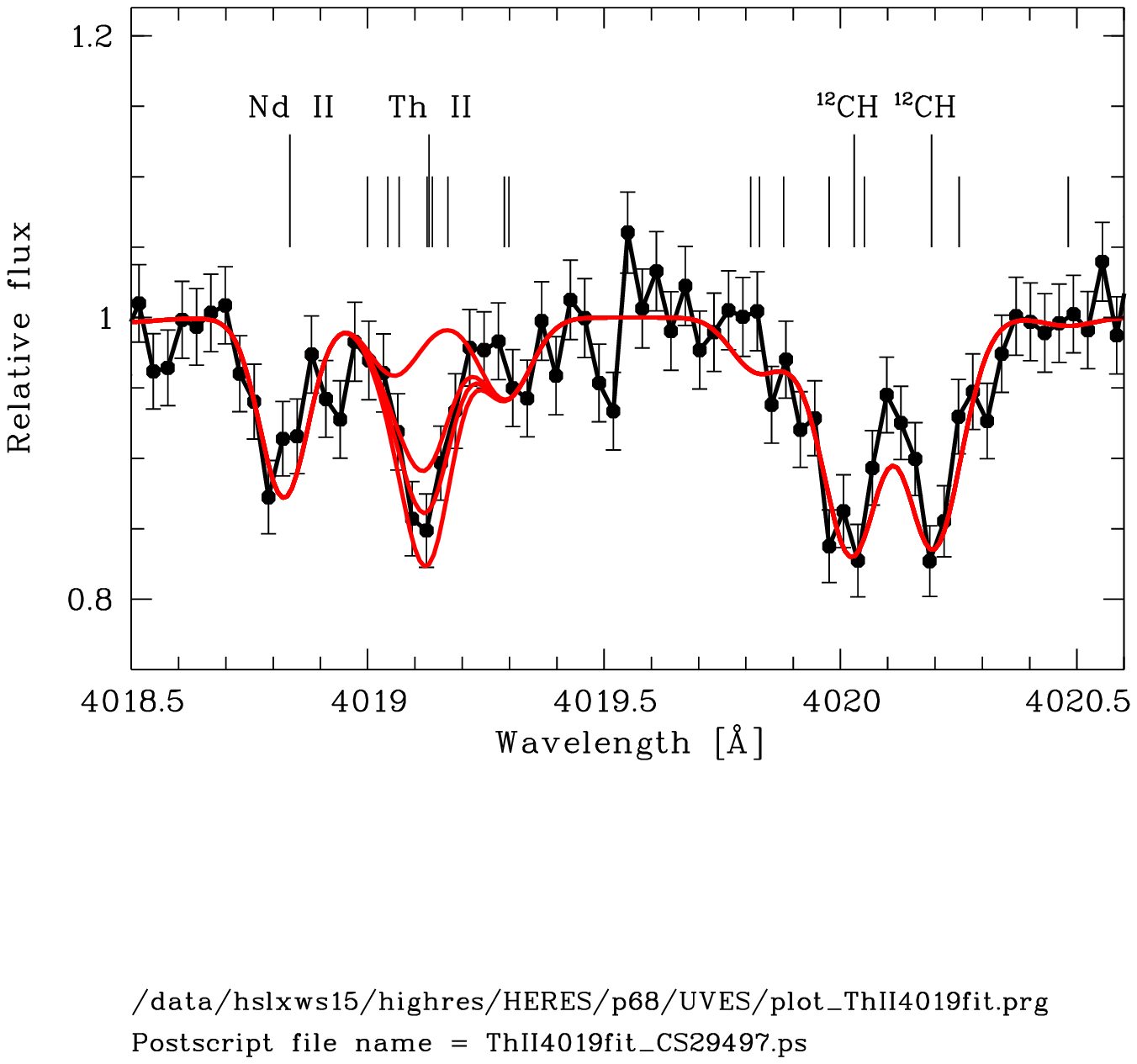, clip=, width=14cm, 
            bbllx=68, bblly=335, bburx=457, bbury=598}
    \caption{\label{Fig:Th4019_synthesis} Spectrum synthesis of the
      spectral region around \ion{Th}{ii} 4019\,{\AA} in {\cs}. The 
      positions of lines that have been taken into account are indicated.
      The observed spectrum is shown as black line; error bars are the
      1\,$\sigma$ noise level per pixel. The grey lines are spectrum
      synthesis results for $\log \epsilon (\mbox{Th}) = -9.99$, $-1.11$, 
      $-0.96$ (best fit), and $-0.81$.}
  \end{center}
\end{figure*}

We detected the \ion{Th}{ii} $4019.129$\,{\AA} line in the UVES snapshot
spectrum of {\cs}, and carried out a spectrum synthesis of the spectral region
around this line (see Figure \ref{Fig:Th4019_synthesis}). It is blended with
lines of other species. In particular, in metal-poor, r-process enhanced stars
the \ion{Ce}{ii} line at $4019.057$\,{\AA} \citep{Snedenetal:1996}, and a
\element[][13]{CH} line at $4019.000$\,{\AA} \citep{Norrisetal:1997b} have to
be taken into account. We used in our spectrum synthesis calculation the line
list and atomic data of \cite{Johnson/Bolte:2001} for this spectral region,
supplemented by a few additional, weak lines identified with the Vienna Atomic
Line Database\footnote{\texttt{http://www.astro.uu.se/$\sim$vald/}}
\cite[VALD;][]{VALD2a,VALD2b}. Whenever the atomic data retrieved form VALD
disagreed with those listed in Johnson \& Bolte, we adopted the former.  
This was the case only for very few lines. 

The most important case is the \ion{Ce}{ii} $4019.057$\,{\AA} line, for which
we adopt the VALD value of $\log gf = -0.213$.  This is $0.306$\,dex lower
than that used by Johnson \& Bolte, which in turn goes back to
\citet{Snedenetal:1996}. The latter authors artificially increased the $\log
gf$ value of this line by 0.3\,dex to improve their fit to the spectrum of
CS~22892-052. However, since \citet{Norrisetal:1997b} have shown that at least
part of the missing absorption on the blue side of the Th line is due to
\element[][13]{CH}, we do not use the scaled $\log gf$ value for the Ce line,
but that listed in VALD.

For the \ion{Th}{ii} $4019.129$\,{\AA} line we use $\log gf = -0.228$
\citep{Nilssonetal:2002}, which is $0.042$\,dex higher than the value adopted
by Johnson \& Bolte, and $0.423$\,dex higher than that listed in VALD. For the
partition functions of thorium we interpolate between the values listed in
Table 3 of \citet{Morelletal:1992}, which are based on calculations provided
by Holweger. In the temperature range covered by our stars, the partition
functions for \ion{Th}{ii} are higher by a factor of $3.6$--$4.4$ than those
listed in \citet{Irwin:1981}, translating to abundance differences of the
order of $0.5$\,dex.

For most of the elements which have lines in the relevant spectral region,
abundances were available from clean lines in other wavelength regions. We
adopted these abundances for the spectrum synthesis.  The abundances of
elements which we could not determine from lines in other wavelength regions
(e.g., Co, Sc) were set to the solar abundance minus $2.64$\,dex (i.e., the
abundances were scaled to the [Fe/H] of {\cs}).

We first used $\element[][12]{C}/\element[][13]{C} = 10$ in our spectrum
synthesis, as constrained from the absence of \element[][13]{CH} features in
other wavelength regions. However, it was found that the fit could be
considerably improved when it was assumed that \emph{all} carbon in {\cs} is
present in the form of \element[][12]{C}, i.e.,
$\element[][12]{C}/\element[][13]{C} = \infty$. Hence we adopted this value
for our final fit.

\begin{figure*}[htbp]
  \leavevmode
  \begin{center}
    \epsfig{file=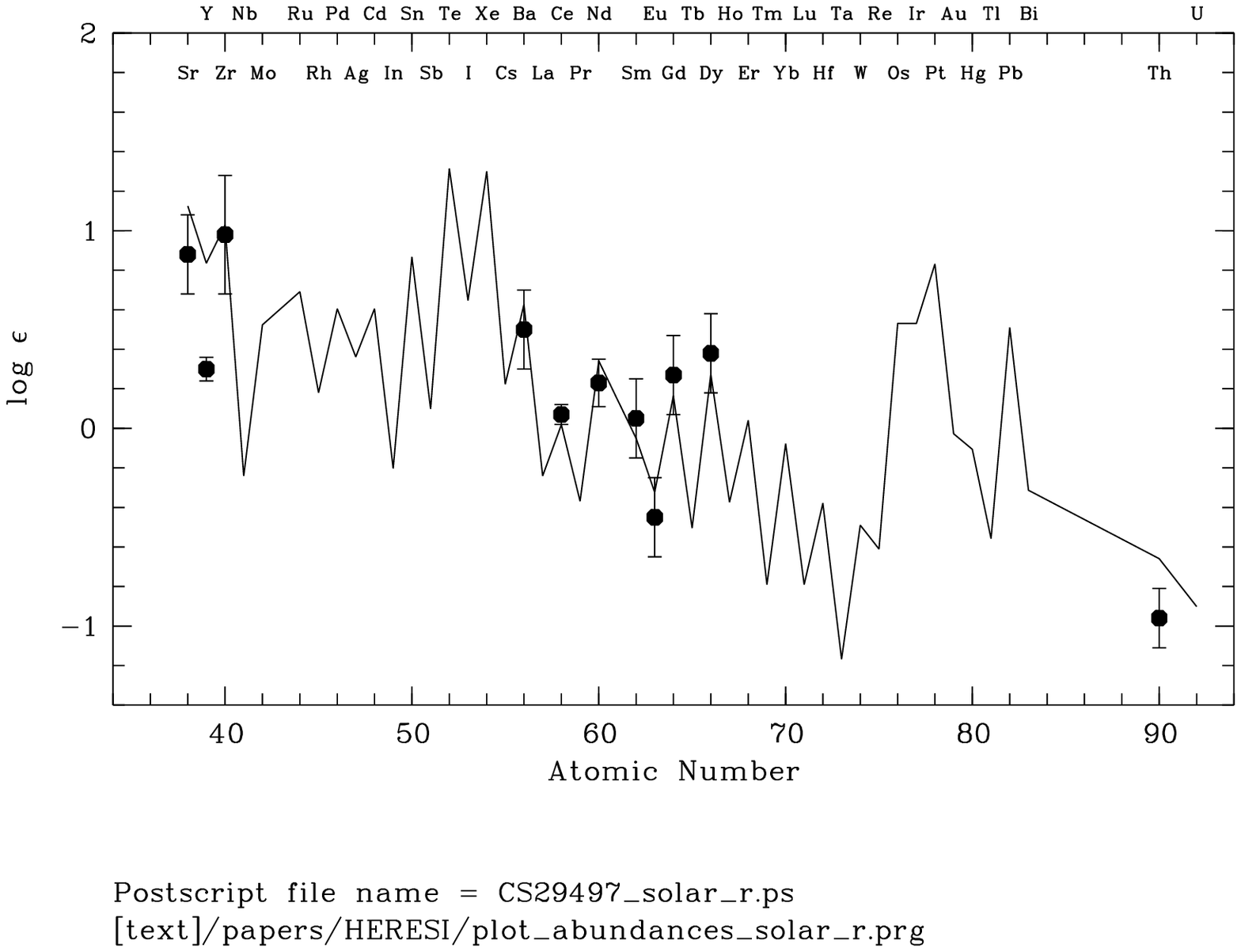, clip=, width=14cm, 
            bbllx=55, bblly=422, bburx=525, bbury=721}
    \caption{\label{Fig:CS29497_solar_r} Abundances of {\cs} compared to the 
      solar r-process abundance pattern, scaled down by 0.83\,dex to match the 
      abundances of Ba to Dy in {\cs}.}
  \end{center}
\end{figure*}

The result is $\log\epsilon (\mbox{Th}) = -0.96$\,dex. We estimate the
uncertainty to be of the order of $0.15$\,dex, which includes uncertainties
introduced by the fit procedure, continuum placement, and the choice of
stellar parameters.

\subsection{Non-LTE effects}

Non-local thermodynamical equilibrium (non-LTE) line formation calculations
for {\cs} have been carried out for \ion{Ba}{ii}, \ion{Eu}{ii}, and
\ion{Sr}{ii} using the methods described in \citet{Mashonkinaetal:1999},
\citet{Mashonkina/Gehren:2000}, and \citet{Mashonkina/Gehren:2001},
respectively. A modified version of the code NONLTE3, based on the complete
linearization method as described by \citet{Auer/Heasley:1976}, has been used
\citep{Kampetal:2004}.

In cool giants, \ion{Ba}{ii}, \ion{Eu}{ii} and \ion{Sr}{ii} are all majority
species.
Therefore, departures from LTE level populations of these species are mainly
caused by radiative bound-bound transitions.  Non-LTE effects strengthen the
\ion{Ba}{ii} and \ion{Sr}{ii} lines compared with the LTE case resulting in
negative non-LTE abundance correction $\Delta_{\mbox{\tiny NLTE}} = \log
\varepsilon_{\mbox{\tiny NLTE}} - \log \varepsilon_{\mbox{\tiny LTE}}$ and
weaken the \ion{Eu}{ii} lines resulting in $\Delta_{\mbox{\tiny NLTE}}$ of
opposite sign. According to our calculations for the investigated spectral
lines $\Delta_{\mbox{\tiny NLTE}} = -0.09$\,dex (\ion{Ba}{ii} $4554$\,{\AA}),
$-0.23$\,dex (\ion{Ba}{ii} $4934$\,{\AA}), $+0.03$\,dex (\ion{Eu}{ii}
$4129$\,{\AA}) and $-0.12$\,dex (\ion{Sr}{ii} $4215$\,{\AA}).

\begin{table}
  \centering
  \caption{\label{Tab:CS29497_QW} Abundances of {\cs} compared with
    predictions of the model of Qian \& Wasserburg.
    }
  \begin{tabular}{lrrr}\hline
    \rule{0.0ex}{2.3ex} & \multicolumn{2}{c}{$\log \epsilon$} \\
    \rb{Element} & \multicolumn{1}{c}{Predicted} & 
    \multicolumn{1}{c}{Observed} \\\hline
    O  & $ 6.82$ & $ $ \rule{0.0ex}{2.3ex}\\
    Na & $ 3.94$ & $ $ \\
    Mg & $ 5.40$ & $ 5.21\pm 0.14$ \\
    Al & $ 3.57$ & $ $ \\
    Si & $ 5.33$ & $ $ \\
    Ca & $ 4.10$ & $ 4.04\pm 0.17$ \\
    Sc & $ 0.57$ & $ $ \\
    Ti & $ 2.75$ & $ 2.67\pm 0.15$ \\
    V  & $ 1.65$ & $ 1.37\pm 0.20$ \\
    Cr & $ 2.84$ & $ $ \\
    Mn & $ 2.28$ & $ $ \\
    Co & $ 2.32$ & $ $ \\
    Ni & $ 3.44$ & $ $ \\
    Sr & $ 0.87$ & $ 0.88\pm 0.20$ \\
    Y  & $ 0.06$ & $ 0.30\pm 0.06$ \\
    Zr & $ 0.65$ & $ 0.98\pm 0.20$ \\
    Nb & $-0.53$ & $ $ \\
    Ru & $ 0.48$ & $ $ \\
    Rh & $-0.31$ & $ $ \\
    Pd & $ 0.22$ & $ $ \\
    Ag & $-0.30$ & $ $ \\
    Cd & $ 0.13$ & $ $ \\
    Ba & $ 0.48$ & $ 0.50\pm 0.20$ \\
    La & $-0.19$ & $ $ \\
    Ce & $ 0.01$ & $ 0.07\pm 0.05$ \\
    Pr & $-0.48$ & $ $ \\
    Nd & $ 0.13$ & $ 0.23\pm 0.12$ \\
    Sm & $-0.17$ & $ 0.05\pm 0.20$ \\
    Gd & $ 0.03$ & $ 0.27\pm 0.20$ \\
    Tb & $-0.67$ & $ $ \\
    Dy & $ 0.11$ & $ 0.38\pm 0.20$ \\
    Ho & $-0.50$ & $ $ \\
    Er & $-0.10$ & $ $ \\
    Tm & $-0.90$ & $ $ \\
    Yb & $-0.19$ & $ $ \\
    Lu & $-0.95$ & $ $ \\
    Hf & $-0.58$ & $ $ \\
    Os & $ 0.37$ & $ $ \\
    Ir & $ 0.40$ & $ $ \\
    Pt & $ 0.69$ & $ $ \\
    Au & $-0.17$ & $ $ \\
    Th & $-0.97$ & $-0.96\pm 0.20$ \\
    U  & $-1.74$ & $ $ \\\hline
   \end{tabular}
\end{table}

\section{The abundance pattern of CS~29497--004}\label{Sect:AbundancePattern}

In Figure \ref{Fig:CS29497_solar_r} we show a comparison of the abundance
pattern of {\cs} with the solar r-process pattern, as listed in Table 5 of
\citet{Burrisetal:2000}, minus 0.83\,dex. This is the average difference
between the abundances of Ba to Dy observed in {\cs} and the solar r-process
pattern. We use the LTE abundances here, because NLTE calculations are not
available to us for all elements for which we have measured abundances. The
abundances of the elements $56 \le Z \le 66$ in {\cs} match the scaled solar
r-process pattern well, while Th is underabundant relative to that pattern by
0.3\,dex, most likely due to radioactive decay.

Y is also underabundant, by about $0.5$\,dex relative to the solar r-process
pattern derived by \citet{Burrisetal:2000}. \citet{Hilletal:2002} investigated
the difference of the solar r-process patterns as derived by
\citet{Burrisetal:2000} and \citet{Arlandinietal:1999}. For Y, these patterns
differ by $\sim 0.5$\,dex. \citet{Hilletal:2002} also noted that a better
match with the observed abundance patterns of CS~22892--052 and CS~31082--001
is reached if the solar r-process pattern of \citet{Arlandinietal:1999} are
used. We hence conclude that the underabundance of Y relative to the scaled
solar r-process pattern we use can not be attributed to a different 
nucleosynthesis history of this element, but is most likely due uncertainties
in the de-composition of the solar abundance pattern into s- and r-process
components. 

\citet{QW01a,QW01b,QW02} proposed a phenomenological model to explain the
abundances of a wide range of elements in metal-poor stars. This model assumes
that in general the abundance of an element E in a metal-poor star is given by
\begin{equation}\label{Eq:QWmodel}
10^{\log\epsilon({\rm E})}=10^{\log\epsilon_P({\rm E})} 
+ n_H\times 10^{\log\epsilon_H({\rm E})} 
+ n_L\times 10^{\log\epsilon_L({\rm E})},
\end{equation}
where $\log\epsilon_P({\rm E})$ represents the prompt ($P$) inventory of E in
the interstellar medium (ISM) contributed by an early generation of stars, and
$\log\epsilon_H({\rm E})$ and $\log\epsilon_L({\rm E})$ represent the yields
of E for two hypothesized ($H$ and $L$) types of core-collapse supernovae that
provide the subsequent enrichment of the ISM.  The quantities $n_H$ and $n_L$
in equation (\ref{Eq:QWmodel}) correspond to the ``numbers'' of $H$ and $L$
events that contributed to the abundances in the ISM from which the star
formed. For {\cs} the model gives $n_H=107$ and $n_L =0.3$ if the observed
abundances of Eu and Fe are taken to be $\log\epsilon({\rm Eu})=-0.45$ and
$\log\epsilon({\rm Fe})=4.81$. Using these values of $n_H$ and $n_L$ and the
$\log\epsilon_P$, $\log\epsilon_H$, and $\log\epsilon_L$ values given in Table
3 of \citet{QW01b} and Table~1 of \citet{QW02}, the abundances of all the
other elements from O to U are calculated from equation (\ref{Eq:QWmodel}) and
presented in Table \ref{Tab:CS29497_QW} (an age of 13\,Gyr is used to
calculate the abundances of Th and U). It can be seen that the model results
are in accord with all the data covering the other elements from Mg to Th to
within $\sim 0.3$ dex.  We note that the abundances of C and Zn cannot be
calculated from equation (\ref{Eq:QWmodel}) due to incomplete knowledge of the
input parameters, especially the relevant $\log\epsilon_L$ values.

In the above model, the values of $n_H = 107$ and $n_L = 0.3$ require that for
{\cs}, the abundances of the elements from O to Ni be dominated by the $P$
inventory and those of the elements from Sr to U by contributions from the $H$
events. Further, the extremely high value of $n_H = 107$ can not be attributed
to enrichment of the ISM but would indicate that the abundances of the
$r$-process elements in CS 29497--004 are due to the contamination of this
star's surface by a previous $H$ event associated with a binary companion. The
question of whether this star is in a binary as evidenced by periodic shift in
radial velocity or was in a binary as evidenced by high proper motion should
be investigated. Thus, {\cs} belongs to the same class of metal-poor stars
with extremely enhanced $r$-process abundances as CS~22892--052 and
CS~31082--001 \citep[see discussion in][]{QW01a}.

We note that the low value of $n_L=0.3$ calculated for {\cs} based on the
model may indicate no $L$ event contributions to this star. Instead, the
abundances of the elements from O to Zn in this star may simply reflect a
possible range of the $P$ inventory due to variations in the mixing of the
contributions from the early generation of stars with the ISM.  In this case,
the abundances of the elements from O to Zn are given by the relevant
$\log\epsilon_P$ values corrected for a dilution factor corresponding to a
shift of $\log\epsilon({\rm Fe})-\log\epsilon_P({\rm Fe})=0.3$ dex. The
results obtained this way for the elements from O to Ni are essentially the
same as those given in Table \ref{Tab:CS29497_QW}. Further, the abundances of
Cu and Zn are calculated to be $\log\epsilon({\rm Cu})=0.83$ and
$\log\epsilon({\rm Zn})=2.16$. The result for Zn is in excellent agreement
with the data.



\section{Conclusions and outlook}

With the discovery of two new r-II stars and one new r-I star, we have
demonstrated that it is indeed feasible to identify such stars by means of
snapshot spectroscopy. We use this concept in a large-scale observational
program -- the HERES project -- which is currently being carried out with
VLT/UVES.

Our pilot project sample contains 23 previously unobserved stars with
$\mbox{[Fe/H]}<-2.5$, and we found two new r-II stars. This exceeds our
previous estimate of the relative frequency of r-II stars of $\sim 3$\,\%. The
full set of 369 HERES stars will allow to determine the relative frequency of
r-II and also r-I stars more accurately.

We measured the abundances of 18 elements in {\cs}, including 11
neutron-capture elements.  The snapshot spectrum we used for our first
analysis was obtained with the aim of identifying r-I and r-II stars, and not
for age determinations. The limited quality of the spectrum results in an
error\footnote{The error analysis procedure described in the Appendix of
Paper~II was employed.} of $0.24$\,dex for $\log \epsilon
(\mbox{Th}/\mbox{Eu})$, and therefore it was not possible to derive a
meaningful age for {\cs} using the Th/Eu chronometer. Furthermore, as we
discussed in the introduction, it is unclear if Th/Eu is a useful chronometer
at all.

It is likely that many more neutron-capture elements can be measured with
higher resolution and $S/N$ spectra. Such spectra have already been obtained
with VLT/UVES. The low carbon abundance ($\log \epsilon (\mbox{C}) = 5.86$)
and high level of enrichment with r-process elements also make {\cs} a good
candidate for an attempt to detect uranium.

We measure $\mbox{[Fe/H]}=-2.64\pm 0.12$ for {\cs}, suggesting that this star
is significantly more metal-rich than the three previously known r-II stars,
CS~22892--052, CS~31082--001 and CS~22183--031 which all have
$\mbox{[Fe/H]}\sim -3.0$. This would have important consequences for future
searches for r-II stars, because it would mean that one should focus on the
metallicity regime $\mbox{[Fe/H]}\la -2.5$.  However, for our analysis of
{\cs} we adopt a temperature scale which is slightly higher than that used in
other studies \citep[see][for a discussion]{KeckpaperI}. The homogenous
analysis of CS~31082--001, {\cs}, and CS~22892--052 presented in Paper~II in
fact indicates that {\cs} has a metallicity in between that of CS~31082--001
and CS~22892--052 (see Table 1 of Paper~II).

In order to determine the boundaries of the metallicity range in which r-II
stars can be found, we are endeavouring in the HERES effort to explore the
region $-2.0 < \mbox{[Fe/H]}< -2.5$ with a statistically significant and
homogenously analysed sample of stars.

\begin{acknowledgements}

We are grateful to the ESO staff at Paranal and Garching for obtaining the
VLT/UVES observations and reducing the data, respectively. Computations of
tailored MARCS models by M. Mizuno-Wiedner, B. Edvardsson and B. Gustafsson
are acknowledged. We are grateful to K. Eriksson and B. Edvardsson for
technical help with the abundance analysis programs, and B. Gustafsson for
many valuable comments. J. Cohen kindly derived an effective temperature for
CS~29497--004 from broad-band photometry for us. We are indebted to A. Frebel
for careful proofreading of an earlier version of the manuscript. N.C.\
acknowledges financial support through a Henri Chretien International Research
Grant administered by the American Astronomical Society, from Deutsche
Forschungsgemeinschaft under grant CH~214/3-1, and a Marie Curie Fellowship of
the European Community program \emph{Improving Human Research Potential and
the Socio-Economic Knowledge} under contract number
HPMF-CT-2002-01437. T.C.B. acknowledges partial funding for this work from
grants AST~00-98508 and AST~00-98549 awarded by the U.S. National Science
Foundation and from award PHY 02-16783, Physics Frontiers Center/Joint
Institute for Nuclear Astrophysics (JINA). P.B.\ acknowledges the support of
the Swedish Research Council. A.J.K. acknowledges support from the
Studienstiftung des deutschen Volkes and the Akademie Leopoldina under grant
BMBF-LPD~9901/8-87. S.R. thanks the Brazilian Institutions FAPESP and 
CNPq for partial financial support.

\end{acknowledgements}

\bibliography{HES,atomdata,instr,modelatm,mphs,ncpublications,ncastro,nonLTE,photometry}
\bibliographystyle{aa}

\end{document}